\documentclass[journal]{IEEEtran}
\usepackage{amssymb,latexsym,amsfonts,amsmath}
\usepackage[a4paper, total={7in, 10.5in}]{geometry}
\usepackage{mathtools,lipsum,cuted}
\usepackage{pbox}
\usepackage{graphicx,color,epstopdf}                     
\usepackage{mathrsfs}                          
\usepackage{upgreek}
\usepackage{caption,tabularx,booktabs}
\usepackage{color, colortbl}
\usepackage{bm}    
\renewcommand\eqref[1]{(\ref{#1})}
\usepackage{subcaption}
\usepackage{slashbox}
\usepackage{diagbox}

\usepackage[sort, numbers]{natbib}
\usepackage{mathrsfs}

\usepackage{multicol}
\usepackage{multirow}
\newcolumntype{L}{>{\arraybackslash}p{0.5cm}}%

\begin{document}

\title{Analysis and Identification of Controller Interaction and Limitation in VSC-HVDC Grid}
\author{\IEEEauthorblockN{Lokesh Dewangan, Himanshu J. Bahirat, \textit{Member, IEEE}}
\thanks{Lokesh Dewangan and Himanshu J. Bahirat are with the Department
of Electrical Engineering, Indian Institute of Technology, Bombay 400076,
India (e-mail: lokeshd@iitb.ac.in; hjbahirat@ee.iitb.ac.in).}
\thanks{}} 

\markboth{}
{Shell \MakeLowercase{\textit{et al.}}: Bare Demo of IEEEtran.cls for IEEE Journals}
\maketitle

\begin{abstract}

With the advancement of the idea of HVDC grid, it becomes imperative to study the interaction of controller and identification of modes of oscillations. The paper presents the complete model of HVDC grid with detailed modeling of controllers, phase lock loops (PLLs), dc grid and the ac systems. An analytical method combining participation factors and eigen-sensitivity is proposed to determine the relative stability of the HVDC grid with respect to parameter variations. The paper also proposes the identification of inter-area and controller interaction modes on the basis of nearest distance between the eigenvalues using the derived sensitivity indices. The critical parameters are varied in a mixed short circuit ratio (SCR) HVDC grid and predictions from the analysis method are confirmed with time domain simulation in PSCAD simulator.

\end{abstract}

\IEEEpeerreviewmaketitle

\section{Introduction}
 
The renewable energy sources have been envisaged as a major source of energy generation for future. These energy sources are often located in remote area and away from the load centers due to environmental, economic and social issues \cite{David}. The integration of renewable energy sources is a challenge because of system stability concerns amongst other problems. Usually, overhead transmission line is used for interconnection of renewable energy sources over short distances. 

For long distance transmission, HVDC line may turn out to be more economical and preferred option. The use of HVDC transmission may facilitate use of underground cables, which can overcome environmental challenges \cite{Acker}. The use of HVDC system eliminates the need of reactive power compensation for long transmission lines and provide enhanced power transfer capability \cite{Arrillaga}. The line commutated converter (LCC) based HVDC system is generally constructed and operated as point to point scheme. These LCC based HVDC systems have large reactive power requirement and voltage polarity reversal is required to reverse direction of power flow. This limits the application of LCC based system to two terminal systems or radially connected  MTDC system \cite{Padiyar}. During last couple of decades, the significant advancements in semiconductor technology has lead to development of voltage source converter (VSC) based HVDC technology. The major benefits with the use of VSC technology are increased of control capability and ease of power reversal, apart from independent active and reactive power control \cite{Yazdani}. The power reversal in VSC based HVDC systems is achieved by changing current direction. Hence, the application of VSC in MTDC system is feasible along with extension to HVDC grids \cite{Arrillaga}. 

The large geographical distance between load and generation centers along with naturally distributed nature of renewable energy sources may indicate feasibility of HVDC grids or MTDC system \cite{Chen}. In addition, asynchronous ac generation could be easily connected to the ac grid. The operation of MTDC system presents several challenges especially related to control of dc bus voltage and power flow. A common practice for point to point HVDC schemes is to control dc voltage at the rectifier terminal and use real power control at the inverter terminal. The renewable energy sources are intermittent in nature therefore converters have to work on both modes for flexible power transmission. 

\begin{figure*}[h]
\centering
    \includegraphics[scale=0.4]{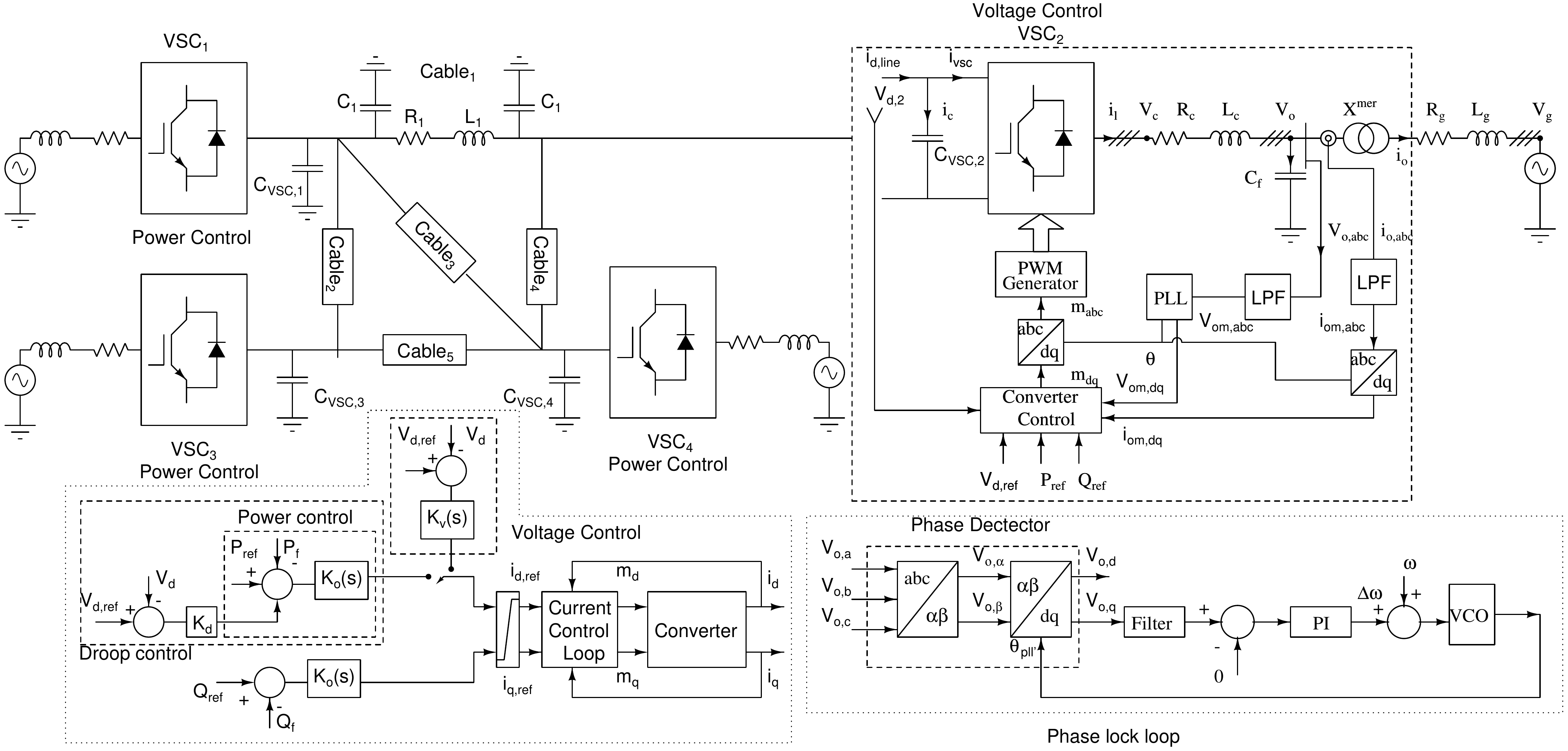}
    \caption{Single line diagram of MTDC system and its control}
    \label{complet}
    \vspace{-0.5cm}
\end{figure*}

A large body of research has been done on the control related issue of MTDC system \cite{Prieto} \cite{Stamatiou} and \cite{Lokesh}. In all control schemes, the primary level controllers are simple PI controllers. These controllers are usually designed and analyzed using static model and studied in  time domain with a convenient system parameters. A numerical approach to study the stability of a VSC by varying the PLL gains and SCR is presented \cite{Gole}. In \cite{Gole}, the studies were conducted with fixed  parameters of outer loop controller. In \cite{Arani}, the parameter variation of outer loop controller is also considered for VSC working in both operating modes and proposed the use of robust or advance controller for VSC connected to a wind farm. In \cite{Zhang}, authors propose replacement of PLL by power synchronization control for VSC connected to weak grid. This however results in increased controller complexity. The \cite{Gole}-\cite{Zhang} are noticed to have one converter or point to point HVDC system.

 The dc grid is represented by $\pi$ model of the transmission line to avoid complexity in small signal model. In \cite{Chaudhuri}, dc network is implemnted using casecaded $\pi$ model and analysis is done for strong grid. The PCC voltage is assumed constant, hence the implementaion of power control loop is not suitable for weak grid. The stability analysis of dc grid has become a wider research area due to increasing possibility of dc resonance as the number of nodes and branches increase \cite{Mura}. These dc resonance frequencies become dominant as the length of line is increased, this also limits the bandwidth of direct voltage control (DVC) \cite{Gustavo}. The higher bandwidth of the DVC limits the maximum power transfer capability and low bandwidth leads to deterioration of system performance. Apart from the parameters of dc grid and ac system, the operating points of VSC also affect the stability of the system \cite{Lokesh1}. For stability assessment, a small signal model of the complete MTDC is required. In \cite{Majumder}, modeling of MTDC grid, the outer loop is implemented using direct expression of power however it is applicable for strong grid. In \cite{Giddani} and \cite{Beerten}, the modelling of VSC based MTDC system is presented, in which feedforward loop is not included and damping controller is added to avoid power oscillation. In \citep{Beerten2}, the interaction between controller for radial network is discussed. The terminal voltage of converter is calculated by multiplication and division of fixed dc voltage, which eliminates the influence of dc dynamics on ac side. The impact of dc grid is a result of dc droop control. In \cite{Shah}, Interaction between grid side converter of wind turbine and HVDC converter interaction is presented. In \cite{Badrkhani}, the interaction between HVDC converter and AC dynamics is presented in time domain simulation. In \cite{Wenjuan}, the interaction between MTDC system and ac power systems is discussed, however this method is applicable only if one dominant pair of each system is near to each other. There is still research gap on controller interaction of HVDC grid when converters are connected to ac systems with mix of SCR values.
      
This paper aims to address the research gap to HVDC grid with multiple SCR ratio by first improving the model of dc grid and detailed dynamic model of HVDC grid. The paper uses the developed small signal model to determine the modes and participation factor for various states on the system operation. The mode participation factor method is further augmented with parametric sensitivity analysis in other to identify local and inter-area modes along with the identification of controller interaction. The method is further demonstrated to be suitable for easily identifying and specifying the stability margin of the HVDC grid with parameter variation. Finally, the prediction and identified limits are confirmed with detailed time domain simulation using PSCAD.    

The organization of the paper is as follows: Section II discusses the combined modeling of the ac and dc systems. The linearized model of the ac-dc system and their augmentation is also presented. In section III, mode participation factor and augmentation of sensitivity is presented to determine the local and inter-area mode along with identification of controller interaction. The validation of derived condition with simulation results    is presented in section IV.  
      
\section{Modeling of HVDC grid} \label{model}
Fig. \ref{complet} shows a four terminal MTDC system with four VSCs connected to four different ac systems with five dc transmission lines interconnecting them. The MTDC system consists of ac system, dc grid, VSCs and its control. In Fig. \ref{complet}, a detailed representation of converter station is shown at terminal-2 (T2) of MTDC system.

\vspace{-0.3cm}
\subsection{Ac system modeling}
Fig. \ref{complet} shows a VSC at T2 connected to an ac grid through filters and transformer along with its control. A PLL is used to estimate frequency of the grid, which is also used for decoupling control design. The reference and measured values are input to the converter control to control power and ac voltage. Further, the modulating signal in $dq$ frame $m_{dq}$ is generated by the controller and converted to $m_{abc}$ using inverse Park's transformation. The $m_{abc}$ values are then used for firing pulse generation with the sinusoidal pulse width modulation (SPWM) technique. The dynamic equations for VSC on the ac side of the converter are given by (\ref{a})
\begin{equation}\label{a}
  \left.
  \begin{array}{ll}
    L_c\frac{d}{dt}i_{l,abc} = -R_ci_{l,abc}+V_{c,abc}-V_{o,abc} \\ [0.3cm]
    L_g\frac{d}{dt}i_{o,abc} = -R_gi_{o,abc}+V_{o,abc}-V_{g,abc} \\ [0.3cm]
    C_f\frac{d}{dt}V_{o,abc} = i_{l,abc}-i_{o,abc}
  \end{array}
  \right\}
\end{equation}
where $R_c$ and $L_c$ are the aggregated resistance and inductance of the phase reactor/filter, respectively; $R_g$ and $L_g$ are the grid resistance and inductance, respectively; a capacitor $C_f$ at the point of common coupling (PCC) is used to filter out higher order harmonics. The systems of equations in (\ref{a}) is transformed into $dq$ domain using synchronously rotating frame of reference that facilitates design of simple PI controller for decoupled control of real and reactive powers. The inner current loops are simple PI controllers that are approximately ten times faster than outer loops \cite{Yazdani}. A voltage feed-forward term is included to avoid the starting transient and noise interference.

\vspace{-0.3cm}
\subsection{Converter and controller model} The time average model of the VSC is considered for small signal analysis of MTDC system. The modeling of conventional vector control which consists of PLL, decoupled inner current control and outer control loop is discussed in details by the authors in \cite{Lokesh1}. It is observed that modeling of VSC generally assumes multiplication and division by constant dc voltage. This eliminates the influence of the dc voltage dynamics on the modulation index calculations. Thus, the dc dynamics do not appear to influence the ac dynamics. In this paper, measured dc voltage is used, with appropriate model of the filter, since it reflects the changes on the dc side to the ac side.. The algebraic equation of converter ac terminal voltage is given by (\ref{Vcd}) 
\begin{equation}\label{Vcd}
  \begin{array}{ll}
    V_{c,d} = \frac{V_{dc}}{V_{dc,f}}(K_P^i(i_{d,ref}-i_{ld})+K_I^i\gamma_{ld}+V_{o,d}-\omega_{pll}L_ci_{l,q}) \\ [0.3cm]
    V_{c,q} = \frac{V_{dc}}{V_{dc,f}}(K_P^i(i_{q,ref}-i_{lq})+K_I^i\gamma_{lq}+V_{o,q}+\omega_{pll}L_ci_{l,d}) 
  \end{array}
\end{equation}
where $V_{dc}$ and $V_{dc,f}$ are dc voltage and filtered dc voltage respectively.

\vspace{-0.3cm}
\subsection{Modeling of dc system}
The dc system of four terminal MTDC of Fig. \ref{complet} is considered to be designed with 300 kV XLPE cables connecting the terminals. The dc cables are represented by $\pi$ equivalent model. The equivalent capacitance ($C_{eqn}$) is obtained by combining dc bus capacitance with capacitance of the cables, where $n$ represents the terminal at which combination is done. Equation (\ref{k}) is obtained by applying KCL at each terminal. The summation of current is obtained  since several cables are connected at each terminal.
\begin{align} \label{k}
C_{eq_n}\frac{dV_{dc_n}}{dt} =& \frac{P_{dc_n}}{V_{dc_n}}\pm\sum_{m=1, \hspace{0.05cm} m \neq n}^{M}i_{dc_{(nm)}},  \\
L_{dc}\frac{di_{dc_{(nm)}}}{dt} =& V_{dc_n} - V_{dc_m} -R_{dc}i_{dc_{(n,m)}}, \hspace{0.2cm} \forall m \neq n,
\end{align} 
where $n$ and $m$ are index of terminals. $\hspace{0.05cm} n = 1,2,.N; m = 1, 2,.M.$ $V_{dc_n}$ and $i_{dc_{(nm)}}$ are the $n^{th}$ bus voltage and line current between $n^{th}$ and $m^{th}$ bus, respectively; $P_{dc_n}$ represents power being exchanged by $n^{th}$ VSC. Equation  (\ref{k}) is linearized around the operating point. The state space representation of the linearized dc system is given by (\ref{m})
\begin{equation}\label{m}
\begin{bmatrix}
\Delta \dot{V}_{dc_n}\\
\Delta \dot{i}_{dc_{(n,m)}}
\end{bmatrix}
=
\begin{bmatrix}
F_n
\end{bmatrix}
\begin{bmatrix}
\Delta V_{dcn}\\
\Delta i_{dc_{(n,m)}}
\end{bmatrix}
+ 
\begin{bmatrix}
E_n
\end{bmatrix}
\begin{bmatrix}
\Delta P_{dc_n}\\
0
\end{bmatrix},
\end{equation}
where $F_n$ and $E_n$ are state and control input matrices respectively.

\vspace{-0.3cm}
\subsection{State space model of MTDC system}  \label{SSMMTDC}
The state space model of a single terminal then obtained by augmenting the linearized state space models of ac system, converter and dc system and is given by (\ref{n}).   
\begin{equation}\label{n}
  \left.
  \begin{array}{ll}
 \Delta \dot{X_n} &= A_{n}\Delta X_n+ B_n \Delta I_{vsc,n} + E_n \Delta U_n \\ [0.3cm]
 \Delta Y_n &= C_n\Delta X_n+ D_n \Delta U_n,
  \end{array}
  \right\}
\end{equation}
where 
  \newline
 $X_n=\left[
\begin{matrix}
x_1&x_2 & x_3 & x_4 & \gamma_{P} & \gamma_Q & \gamma_{ld} & \gamma_{lq} & \theta_{pll} & \omega
\end{matrix} \right.$

  \hspace{0.5cm}  $ \left.     
     \begin{matrix}
     i_{l,d} & i_{l,q} & V_{o,d} & V_{o,q} & i_{o,d} & i_{o,q} & V_{dc} & x_5
     \end{matrix} 
    \right] $ \vspace{0.3cm}
        
The state variables ($x_1, x_2, x_3, x_4$ and $x_5$) are corresponding to the filters dynamics; $\gamma_P$ and $\gamma_Q$ are states of outer control loop; $\gamma_{ld}$ and $\gamma_{lq}$ are states of inner control loop. Equation (\ref{o}) gives the state space model for the entire MTDC system obtained by augmentation. 

\begin{tiny}
\begin{equation} \label{o}
\begin{bmatrix}
\Delta \dot{X_1}\\
\Delta \dot{X_2}\\
\dots\\
\Delta \dot{X_N}\\
\Delta \dot{i_{dc_{(n,m)}}}
\end{bmatrix}
=
\begin{bmatrix}
A_{1} & 0 & 0 & 0 & B_1\\
0 & A_{2} & 0 & 0 & B_2\\
0 & 0  & \dots & 0 & \dots\\
0 & 0 & 0 & A_{N} & B_N\\
F_1 & F_2 & \dots & F_M & G
\end{bmatrix}
\begin{bmatrix}
\Delta X_1\\
\Delta X_2\\
\dots\\
\Delta X_N\\
\Delta i_{dc_{(n,m)}}
\end{bmatrix}
+
\begin{bmatrix}
\Delta E_1\\
\Delta E_2\\
\dots\\
\Delta E_N
\end{bmatrix}
\begin{bmatrix}
\Delta U_1\\
\Delta U_2\\
\dots\\
\Delta U_N
\end{bmatrix}
\end{equation}
\end{tiny}
where F and G are the matrices related to the dc system. $A_{n}$ and  $B_{n}$ are state and control matrix of each VSC respectively. In this space-state model, each converter station has 16 states (6 states are controller state, 6 states are ac system dynamics, 4 states are filter dynamics) and dc system has 13 states (4 dc terminal voltages, 4 filtered dc voltages and 5 line currents). The control input to the system is $u=[P_{ref}\hspace{0.2cm}V_{dc,ref}]$. The derived model is validated with non-linear model of HVDC grid implemented in PSCAD.

\vspace{-0.3cm}
\subsection{System description} \label{description}
The MVA rating of each VSCs are 600 MVA and its primary control modes are depicted in the Fig. \ref{complet}. The parameters of the converter station, controllers and dc transmission lines are given in the Table \ref{VSC_par} and \ref{Con_par}. The T2 and T4 are connected to systems having low SCRs, for example: wind farm, are 1.5 and  1.8 respectively. The T2 and T4 are assumed to be in inversion and rectification mode with active and reactive power controllers respectively. Terminals T1 and T3 are working in rectification and inversion mode respectively. T1 is provided voltage controller to regulate the dc voltage of entire grid and T3 is equipped with active and reactive power control. The terminals T1 and T3 are connected to ac grid having SCR of 3.2 and 2.0. According to IEEE 1204-1997, T2 and T3 are classified as very weak and weak grid respectively.

\begin{table}[h]
\caption{Parameters of the VSC and dc grid} \label{VSC_par} 
\begin{center}
\vspace{-0.3cm}
\begin{tabular}{ c|c|c|c }
\hline
 Parameters & Actual value & Parameters & Actual value \\
 \hline
 \hline 
 $MVA$ & 600 $MVA$ & $R_{dc}$ & 0.0121 $\Omega/km$ \\  
 $V_{ac}$ & 300 $kV$ &  $L_{dc}$ & 0.1056 $mH/km$ \\
 $MW$ & 600 $MW$ & $C_{dc}$ & 0.2961 $uF/km$ \\
 $V_{dc}$ & 600 $kV$ & $C_{vsc}$ & 66.66 $uF$ \\
 $C_{f}$ & 3.12 $uF$ & length of cable1 & 120 $km$\\
 $L_c$ & 71.6 $mH$ & length of cable2 & 200 $km$ \\  
 $R_c$ & 0.225 $\Omega$ & length of cable3 & 80 $km$ \\
 $L_g$ & 50-460 $mH$ & length of cable4 & 160 $km$\\
 $R_g$ & 1.975 $\Omega$ & length of cable5 & 160 $km$\\
 \hline
\end{tabular}
\end{center}
\vspace{-0.5cm}
\end{table}

\begin{table}[h]
\caption{Parameters of the Controllers and Filters} \label{Con_par} 
\begin{center}
\vspace{-0.3cm}
\begin{tabular}{ c|c }
\hline
 Controllers and filters & values \\
 \hline
 \hline 
 Inner controller ($K_P^i$, $K_P^i$) & ($L_c/\tau_{i}$, $R_c/\tau_{i}$, $\tau_i = 1ms$) \\  
 PLL controller ($K_P^{pll}$, $K_I^{pll}$) & (10-100, 50-500) \\
 Power controller ($K_P^P$, $K_I^P$) & (0.1, 100) \\
 Voltage controller ($K_P^{dc}$, $K_P^{dc}$) & (2.25, 100) \\
 Voltage filter ($T_{mvd}$, $T_{mvq}$) & (0.02, 0.02) \\
 Current filter ($T_{mid}$, $T_{miq}$) & (0.0012, 0.0012)\\
 \hline
\end{tabular}
\end{center}
\vspace{-0.25cm}
\end{table}

It is usually observed in expansive networks that there exist interaction between regions that are separated by large distance. In addition, it is also observed that there is a local interaction between generators. It is expected that such interaction will exist when HVDC grid is built. Thus, the system model derived in this section will be used in the subsequent section to identify the modes of interaction and define the stability boundaries, which may be dependent on system parameters.   

\section{Analysis and identification of interactions} \label{Method}
In a literature, various methods (like Prony, Eigenvalue realization and matrix pencil) are used for electromechanical oscillation identification of power system. However, the modes are classified based on state participation \cite{Kundur}. With reference to system model derived in the section, the controller states are the state variables and system parameters form the state matrix. In the power system literature, the state matrix is usually used to identify the eigenvalues and mode of interaction. From the point to point HVDC literature, it is observed that the PLL gains and the corresponding SCR impact the stability. Usually, weak grids are considered and controller design for outer loop is done with a bandwidth which is 5 to 10 times higher than a bandwidth of PLL. It may be possible that during certain operating condition, there might be an interaction between the PLL and controller of an outer loop. This problem of interaction may get worse when HVDC grids are constructed. Especially, when ac systems have different SCR values.

In a case of HVDC grid as well, the eigenvalue analysis can be used to identify the modes. The well-known participation factor analysis could be used to analyze the impact of parameter variation on these modes. However, the participation factor method does not provide any information on the stability margin or degree of interaction between the modes. Thus this paper extends the participation analysis further with the help of sensitivity analysis to define the boundary condition to ensure the stability and also define the method to identify the degree of interaction.

\vspace{-0.3cm}
\subsection{Mode identification} \label{SectPart}
This section summarizes the participation factor method and the eigenvalue analysis. In \cite{Lokesh1}, the variation of PLL gain $K_P^{pll}$ and grid inductance $L_g$ caused huge variation on two dominant eigenvalues. Therefore, it may be concluded that only a few states of the converters are responsible for the interaction. These states may be identified using state participation. For eigenvalue analysis state matrix ($A_{sys}$) from (\ref{o}) is used. The state participation factor corresponding to $A_{sys}$ is given by (\ref{Participation})
\begin{equation} \label{Participation}
\mathbb{P}_{ok} = Y^T_{ok}X_{ko}
\end{equation}
where $X_{ok}$ and $Y_{ok}$ are the left and right eigenvectors for $o^{th}$ state and $k^{th}$ eigenvalue respectively. $\mathbb{P}_{ok}$ shows the participation of $o^{th}$ state on $k^{th}$ eigenvalue. The classification and identification is done on the basis of the magnitude and phase of the participation factors.

The state matrix is made by augmenting the state of various terminals. Hence the participation factor corresponding to VSC is defined by (\ref{Participation}), which combines the impact of the state variables of the given VSC on the particular mode. This facilitates easy identification of the criteria for mode classification. It can be easily seen that the algebraic sum of the participation factor of VSC is one.
\begin{equation} \label{PSUM}
\sum_{n=1}^N \mathbb{P}_{VSC_n} = 1
\end{equation}

Thus if a mode with $\mathbb{P}_{VSC_n}$ greater than 0.3 is considered to be impacted by the $VSC_n$ under consideration. This mode is then defined as the local mode. If the participation factor ($\mathbb{P}_{VSC_n}$) of more than one terminals is greater than 0.3 then the mode is defined as an inter-area mode. The participation factor corresponding to particular frequency may itself be dominated by state variable corresponding to electrical quantity or controller states. If $\mathbb{P}_{VSC_n}$ is dominated by a controller states and mode is identified as local mode then it is termed as local control mode (LCM). On the other hand, if a mode is identified as an inter-area mode then the interaction is termed as inter-area control mode. The condition correspond to local and inter-area control mode (ICM) are given by (\ref{MCLASS})

\vspace{-0.3cm}
\begin{align} \label{MCLASS} \nonumber
 LCM: \mathbb{P}_{\gamma_Pk} \approx \mathbb{P}_{\omega k}, \mathbb{P}_{\gamma_Pk} \approx \mathbb{P}_{\theta_{pll} k}, \\
 \mathbb{P}_{\gamma_Qk} \approx \mathbb{P}_{\omega k}, \mathbb{P}_{\gamma_Qk} \approx \mathbb{P}_{\theta_{pll} k}, \\
 ICM: \mathbb{P}_{X_ok} \approx \mathbb{P}_{X_qk} \hspace{1.2cm} \forall o \neq q \nonumber
\end{align} 

It is observed that for HVDC grids the controller mode may be dominant mode because of fast dynamics.
The real part of the participation is a necessary condition for classification but it does not provide complete information about the nature of the oscillations. If the tendency of the state participation factor is to reduce the absolute real part of the eigenvalue then it makes the system more oscillatory. Whereas it increases the absolute value of the system become more stable. This can be easily observed from eigen sensitivity

\vspace{-0.3cm}  
\subsection{Sensitivity analysis}
In this section, the sensitivity of the HVDC grid is derived and consequently some analytical boundary conditions are obtained for smooth operation of HVDC grid. The eigen sensitivity for matrix $A_{sys}$ corresponding to HVDC grid can be obtained by solving (\ref{SENE})
\begin{equation} \label{SENE}
\frac{\partial \lambda_k}{\partial q} = Y_k^T\frac{\partial A_{sys}}{\partial q}X_k
\end{equation}

If the parameter corresponds to the diagonal element of $A_{sys}$, the eigen sensitivity is equal to the participation factor. In the case of the off-diagonal element, the eigen sensitivity is not directly related to the participation factor. The elements of the state matrix ($A_{sys}$) depend upon the parameters like the controller and PLL gain and the grid inductance. The change in eigenvalues corresponding to changes in the parameter can then be given by (\ref{GS})
\begin{equation} \label{GS}
\Delta \lambda_k = \sum_{j = 1}^{J} Y_k^T\frac{\partial A_{sys}}{\partial q_j}X_k \Delta q_j
\end{equation}

Based on the literature survey and the demonstration from result section \ref{Results}, it is seen the most important parameter impacting the stability are the PLL gain and grid inductance. The grid inductance can also be though as the measure of SCR. With consideration of these parameters, only the derivative matrix is obtained to be a sparx matrix, which facilitates the development of analytical relationships. The sensitivity coefficient of the dominant mode identified by the analysis presented in a section is then obtained to be equal to (\ref{KPLG}) and (\ref{KPLG1}).
\begin{equation} \label{KPLG}
\Delta \lambda_k =  Y_k^T\bigg[\frac{\partial A_{sys}}{\partial K_P^{pll_n}} \Delta K_P^{pll_n}+\frac{\partial A_{sys}}{\partial L_{g_n}} \Delta L_{g_n}\bigg]X_k
\end{equation} 
\begin{equation} \label{KPLG1}
\Delta \lambda_k =  \bigg[\frac{\partial \lambda_k}{\partial K_P^{pll_n}} \Delta K_P^{pll_n}+\frac{\partial \lambda_k}{\partial L_{g_n}} \Delta L_{g_n}\bigg]
\end{equation} 

The further expansion of the sensitivity with respect to $K_P^{pll}$ proportional gain is then given by (\ref{kpll}) and it is seen to be dependent on the value of filter capacitance, the d-axis voltage, capacitance and frequency of the corresponding VSC.
\begin{align} \label{kpll}
\frac{\partial \lambda_k}{\partial K_P^{pll_n}} = l_{(k,\mathbb{N}+10)}\bigg[V_{o,d}r_{(\mathbb{N}+10,k)}-\frac{r_{(\mathbb{N}+12,k)}}{C_f}\\ \nonumber
+\frac{r_{(\mathbb{N}+14,k)}}{C_f}-wr_{(\mathbb{N}+15,k)}+5r_{(\mathbb{N}+16,k)}\bigg]
\end{align}
\begin{footnotesize}
\begin{align}\label{Lg} \nonumber
\frac{\partial \lambda_k}{\partial L_{gn}} = l_{(k,\mathbb{N}+13)}\Bigg[\frac{V_msin(\delta)r_{(\mathbb{N}+9,k)}+R_gr_{(\mathbb{N}+13,k)}+r_{(\mathbb{N}+15,k)}}{L_g^2}\Bigg] \\ 
+l_{(k,\mathbb{N}+14)}\Bigg[\frac{V_mcos(\delta)r_{(\mathbb{N}+9,k)}+R_gr_{(\mathbb{N}+14,k)}+r_{(\mathbb{N}+16,k)}}{L_g^2}\Bigg]
\end{align}
 \vspace{-0.3cm}
\end{footnotesize}

In (\ref{kpll}), $l_{(k,\mathbb{N}+10)}$ and $r_{(\mathbb{N}+10,k)}$ represent the element of the left and right eigenvector corresponding to $k^{th}$ eigenvalue respectively. Here the row index of the element depends upon the serial number of the converter (n). Thus for the T3, the index would be 18(3-1)+10 = 46 element. Based on the analysis it is observed that the sensitivity index (\ref{kpll}) is large invariant when d-axis voltage and the element $r_{(\mathbb{N}+16,k)}$ are changed. Hence reduced index is given by (\ref{kpllr})
\begin{align} \label{kpllr}
\frac{\partial \lambda_k}{\partial K_P^{pll_n}} = l_{(k,\mathbb{N}+10)}\bigg[-\frac{r_{(\mathbb{N}+12,k)}}{C_f}-wr_{(\mathbb{N}+15,k)} +\frac{r_{(\mathbb{N}+14,k)}}{C_f}\bigg]
\end{align}

The new eigenvalue corresponding to change in PLL gain $K_P^{pll}$ and grid inductance $L_{gn}$ can then be obtained.
\begin{equation} \label{NE}
\lambda^{new}_k = \lambda^{old}_k + \frac{\partial \lambda_k}{\partial K_P^{pll_n}} \Delta K_P^{pll_n}+ \frac{\partial \lambda_k}{\partial L_{gn}} \Delta L_{gn}
\end{equation}

The stability index now can be defined as the real part of the eigenvalue and sensitivity coefficient. The condition to ensure stability is thus given by (\ref{cond1})
\begin{equation} \label{cond1}
| Re(\lambda_k^{K_P^{pll} =\mathbb{Z}}) | \leq  |Re \bigg[ \frac{\partial \lambda_k}{\partial K_P^{pll}} \Delta K_P^{pll} \bigg ]_{K_P^{pll} =\mathbb{Z}}|
\end{equation}

The controller interaction and degree of interaction can be further defined by the sensitivity coefficient and distance between the dominant eigenvalues. The distance can basically be obtained by looking the bandwidth of the corresponding controllers. The approximate bandwidth of the controllers are equal to real part of eigenvalues. Therefore distance between two eigenvalues corresponding to controllers must greater than four. The condition to avoid controller interaction is thus given by (\ref{cond2})

\vspace{-0.3cm}
\begin{footnotesize}
\begin{align} \label{cond2} \nonumber
\Bigg [ \bigg(Re(\lambda_k)+Re \bigg[ \frac{\partial \lambda_k}{\partial K_P^{pll}} \bigg ]-Re(\lambda_{\mathbb{K}})-Re \bigg[ \frac{\partial \lambda_{\mathbb{K}}}{\partial K_P^{pll}} \bigg ] \bigg)^2 \\
 +\bigg (Im(\lambda_k)+Im \bigg[ \frac{\partial \lambda_k}{\partial K_P^{pll}} \bigg ]-Im(\lambda_{\mathbb{K}})-Im \bigg[ \frac{\partial \lambda_{\mathbb{K}}}{\partial K_P^{pll}} \bigg ] \bigg)^2 \Bigg]^{1/2} \leq \epsilon
\end{align} 
\vspace{-0.3cm}
\end{footnotesize}

Above two boundary condition is applicable for other parameters variation. Hence, the boundary condition  for grid inductance is not written due to repetition of same equation with different variable. 

\section{Simulation result} \label{Results}
The state matrix corresponding to four terminal HVDC grid shown in Fig. \ref{complet} is constructed using the system parameter given in Table \ref{VSC_par} and \ref{Con_par}. Corresponding to HVDC grid the eigenvalues and participation factor are determined using (\ref{o}) and (\ref{Participation}). A single terminal has 18 states. The overall system as pointed in section \ref{model} has 77 states. Table IV gives the modes for the four terminal HVDC grid along damping factor. A particular mode is considered as dominant mode close to imaginary axis and low damping factor. Thus, the modes on the real axis close to origin may not be dominant mode. The modes-7 to 14 shown in Table IV are not significantly impacted by variation of system and controller parameters.

\begin{table}[h]
\caption{Test Cases for Mode Identification} \label{Cases} 
\centering
\begin{tabular}{ c|c|c|c|c }
\hline
 Terminals & T1 & T2 & T3 & T4 \\
 \hline
 \hline
 Control Strategies & $V_{dc}, Q$ & $P,Q$ & $P,Q$ & $P,Q$ \\[0.25cm]
 \hline 
 Change in OM & Inv & Rec & Inv to Rec & Rec to Inv\\[0.25cm]
 \hline
 Change in SCR & 3.2 & 1.5  & 2  & 1.8 to 1.6 \\[0.25cm]
 \hline
 Change in $K_P^{pll}$ & 10 & 10 & 10 & 10 to 30\\[0.25cm]
 \hline
\end{tabular}
\vspace{-0.3cm}
\end{table} 

\begin{figure}[h]
	\centering
    TABLE IV: Eigenvalues of HVDC grid
    \includegraphics[clip, trim=1.25cm 13cm 4.5cm 2.2cm, width=0.5\textwidth, height=0.5\textwidth]{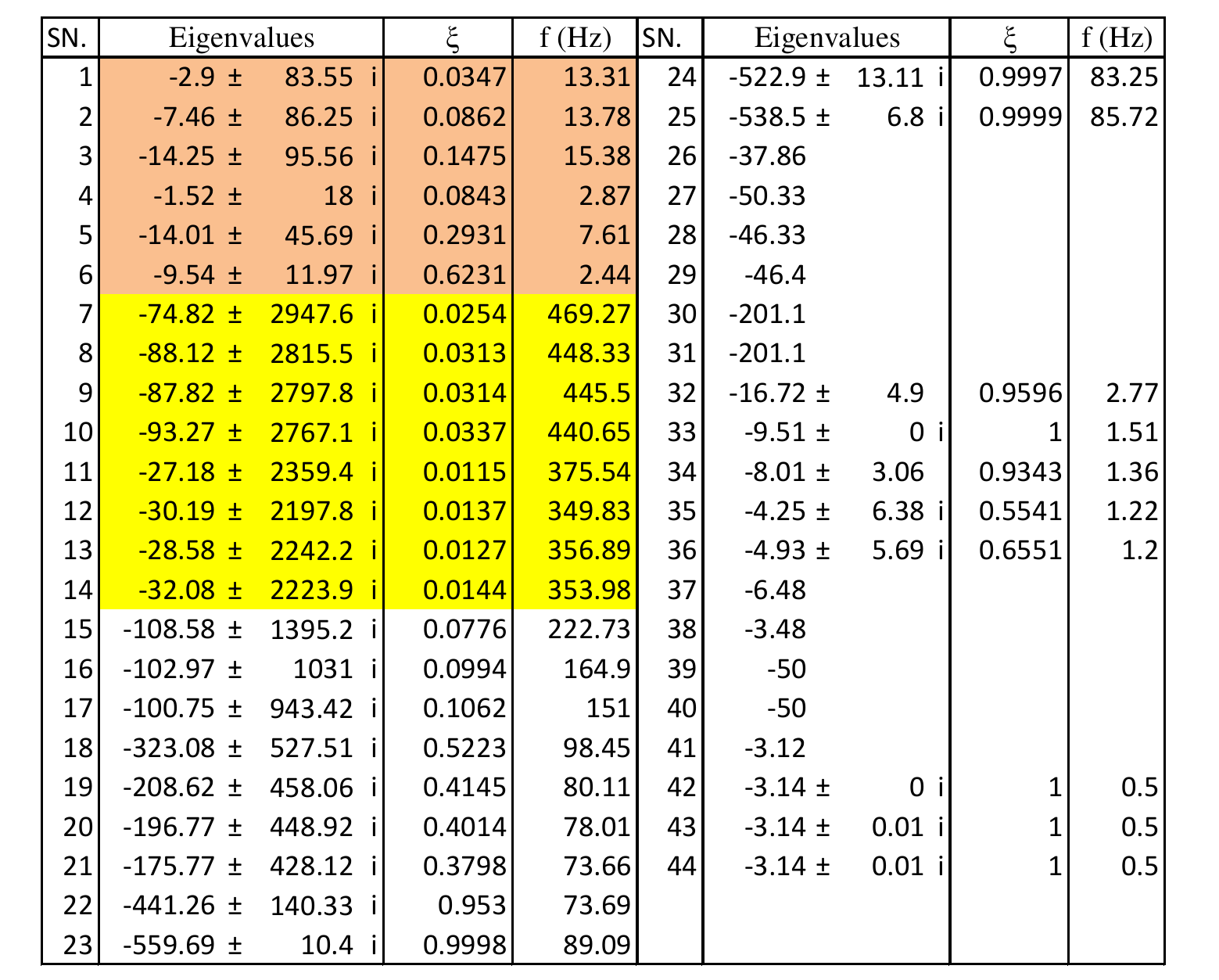}
    \label{Ts}
    \vspace{-0.5cm}
\end{figure}

\begin{figure*}[h]
\centering
    \includegraphics[width=1\textwidth, height=0.20\textwidth]{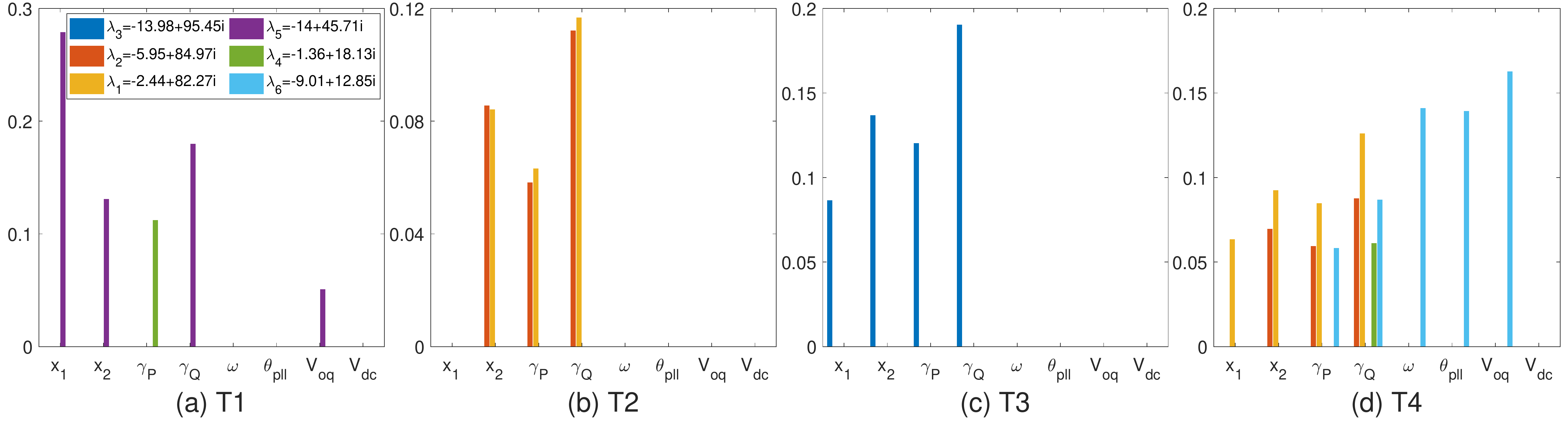}
    \caption{Y-axis: normalized state participation on six dominant eigenvalues, x-axis: states of VSC}
    \label{Ts}
    \vspace{-0.5cm}
\end{figure*}


\begin{figure*}[h]
\centering
    \includegraphics[width=1\textwidth, height=0.20\textwidth]{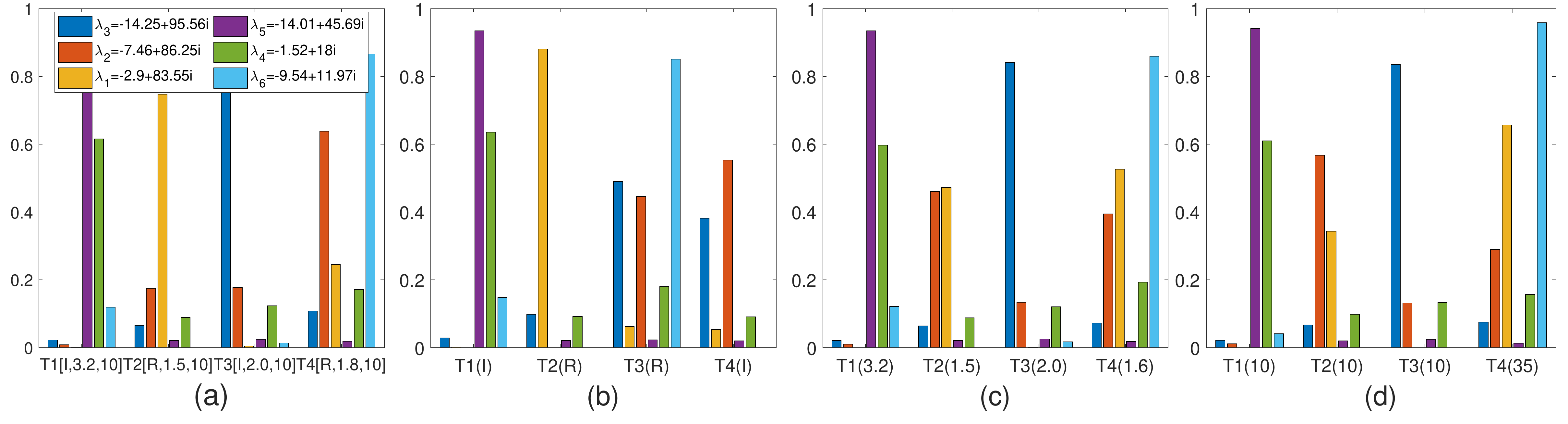}
    \caption{\footnotesize{ Y-axis: normalized terminal participation on six dominant eigenvalues, x-asix: converters, (a) base case (b) change in operating mode (c) change in SCR (d) change in PLL gain}}
    \label{OSP}
    \vspace{-0.5cm}
\end{figure*}

\begin{figure*}[h]
\centering
    \includegraphics[width=1\textwidth, height=0.25\textwidth]{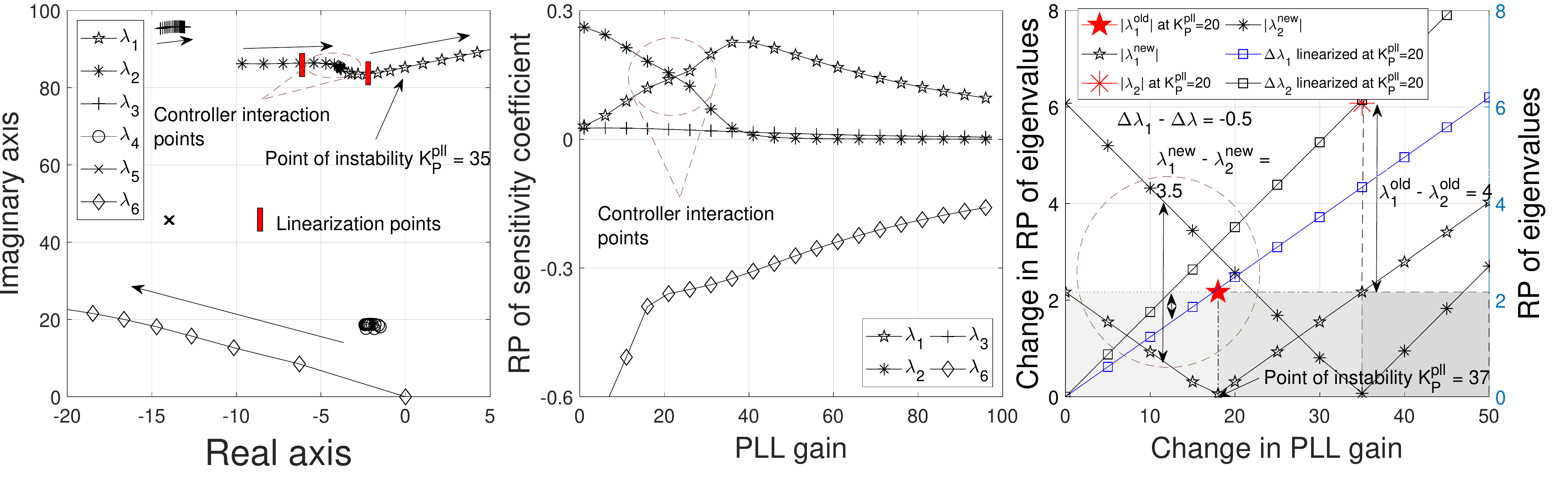}
    \caption{\footnotesize{(a) Root locus of HVDC grid to change in PLL gain (b) Sensitivity of roots with PLL gain changes (c) Conditions of controller interaction and instability of HVDC grid}}
    \label{RootPLL}
    \vspace{-0.5cm}
\end{figure*}

\begin{figure*}[h]
\centering
    \includegraphics[width=1\textwidth, height=0.25\textwidth]{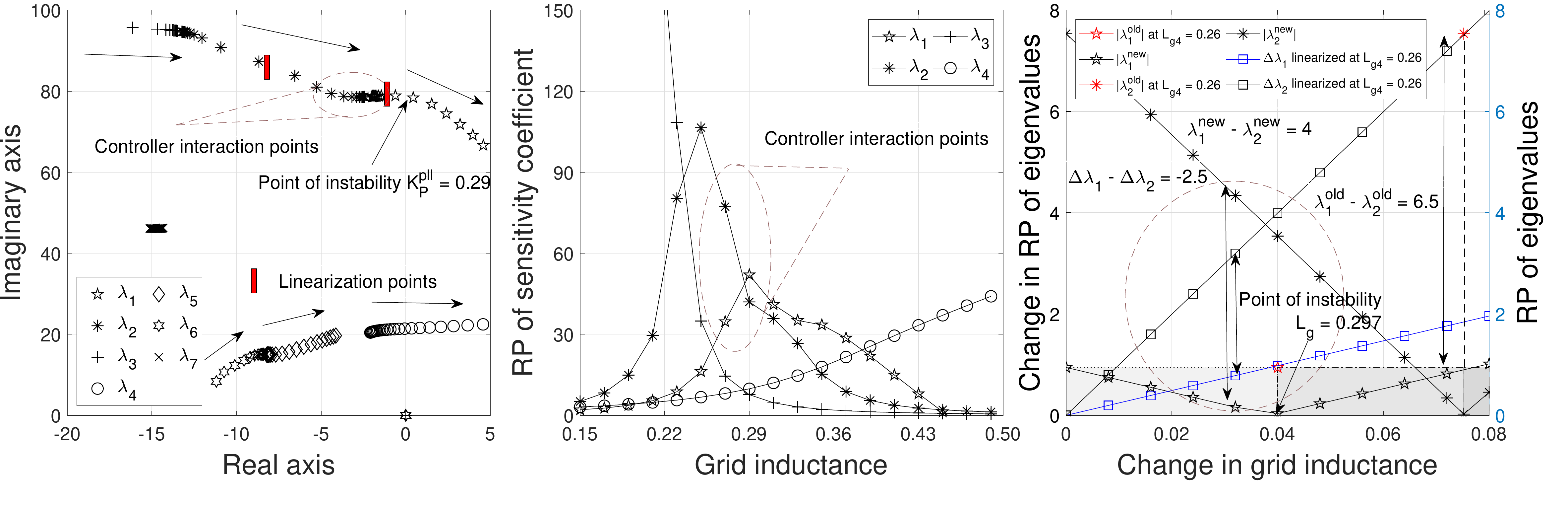}
    \caption{\footnotesize{(a) Root locus of HVDC grid to change in SCR (b) Sensitivity of roots with SCR changes (c) Conditions of controller interaction and instability of HVDC grid}}
    \label{RootSCR}
    \vspace{-0.5cm}
\end{figure*}

\begin{figure}[h]
\centering
    \includegraphics[width=0.5\textwidth, height=0.34\textwidth]{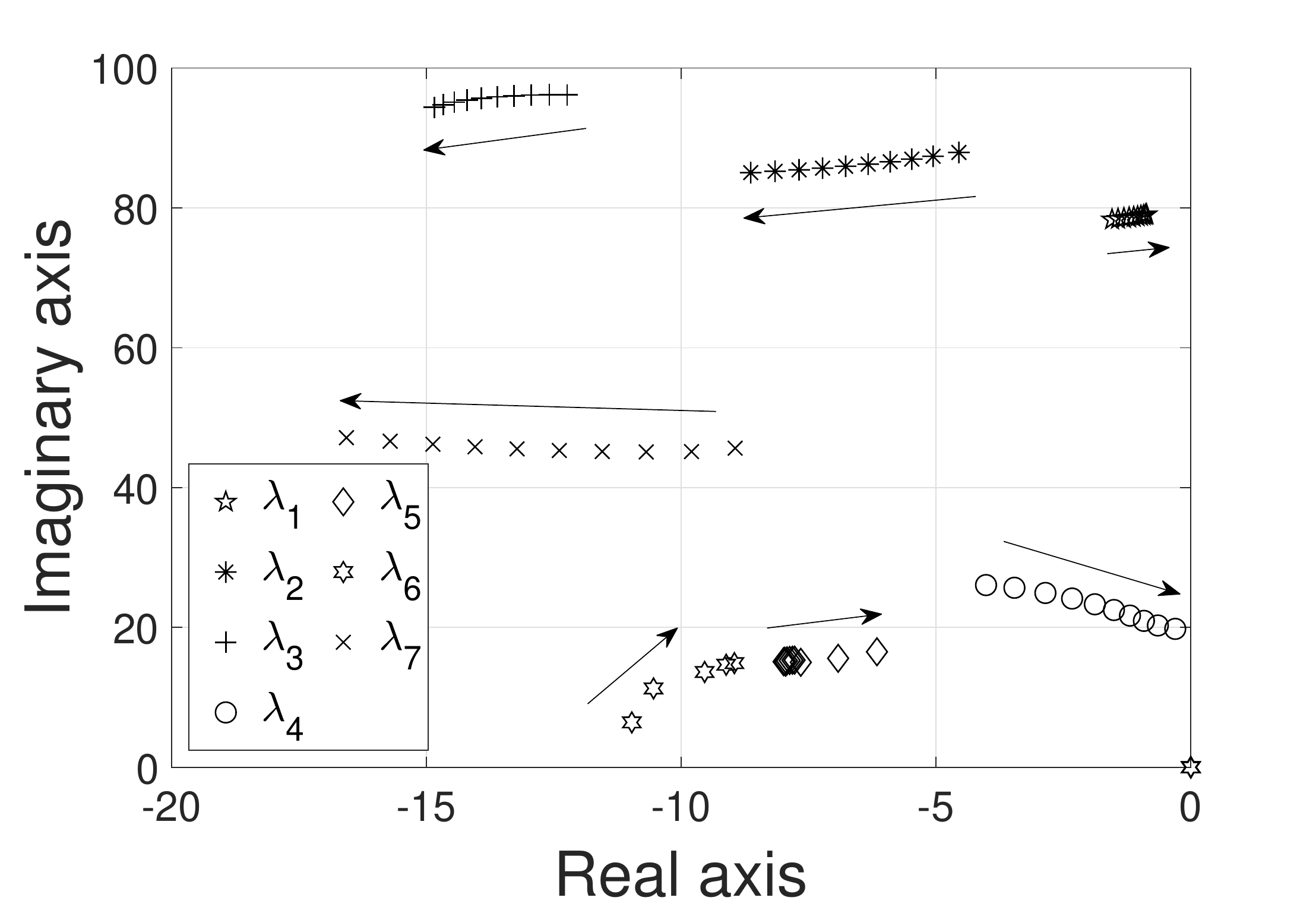}
    \caption{\footnotesize{Root locus of HVDC grid to power changes of T4 }}
    \label{RootLoad}
    \vspace{-0.5cm}
\end{figure}

These modes are not dominant because they lie far from imaginary axis and their movements are not significant to cause instability. The modes-1 to 6 are dominant modes because they are close to imaginary axis and low damping factors. These modes are significantly impacted by variation of system and controller parameters. Fig. \ref{Ts}-\ref{OSP} shows the normalized state participation factor for dominant poles of HVDC grid. As defined in section \ref{Method}, the impact of the states of the VSC terminals is identified first. For given parameters in section \ref{description}, no interaction mode is found therefore Fig. \ref{Ts} is plotted for T4 with SCR of 1.6. From Fig \ref{Ts}a, it can be seen that mode-4 and mode-5 are impacted by states of the T1, From Fig. \ref{Ts}b, T3 impacts mode-3 whereas  T2 and T4 are seen to have greater impact on mode-1, 2 and 6 respectively. The combined impact of the VSC on the dominant mode is given in Fig. \ref{OSP}. Three different conditions are considered to identify controller interaction as given in Table \ref{Cases}.   From Fig. \ref{OSP}a it can be seen that all modes are local modes since each mode is impacted by individual terminal.

On comparing Fig. \ref{OSP}a and Fig. \ref{OSP}b it is noted that the participation indexes depends upon the operating condition of the terminals. When T4 is changed from rectification to inversion operation and vice-versa for T3, the impact on mode-6 is changed from T4 to T3 and mode-2 and mode-3 have partial impact of both terminals. Mode-2 and mode-3 transform into inter-area modes. From Fig. \ref{OSP}c and \ref{OSP}d, the participation factors are seen to significantly impacted by SCR and PLL gain of the converters. The change of parameters also lead to transition of certain local mode to inter-area mode and vice versa. The effect of change in SCR depicted in Fig. \ref{OSP}c shows that the mode-1 and mode-2 transition to become the inter-area modes whereas mode-3, 4, 5 and 6 remain to be a local modes. Fig \ref{OSP}c shows that participation index of T4 increases in mode-1 when SCR is reduced whereas that of T2 reduces. The participation in mode-2 is reduced for T4 and increases for T2. Thus it can be concluded that the operating condition and the SCR have significant impact on the participation factor in a certain modes. It also indicates that the transition of mode is governed by these factors. 

Fig. \ref{OSP}d shows that modes transition to become inter-area mode when PLL gain for T4 is increased to 35 which represent high bandwidth PLL. Mode-1 and mode-2, which were local modes impacted by T2 and T4 respectively, are transform into inter-area modes. Now, mode-1 and mode-2 have equal impact on T2 and T4. It can be concluded that the PLL gain may increase inter-area oscillation. From the above discussion one can not conclude regarding relative stability. Thus Fig. \ref{RootPLL} and \ref{RootSCR} show the root locus and sensitivity plots when PLL gain and SCR are changed respectively.  

In Fig. \ref{RootPLL}a, the root locus of HVDC grid is plotted. Since the bandwidth of the outer controller is usually close to 100 Hz, hence the range of PLL gain ($K_P^{pll}$) of T4 is chosen to be 1 to 100. The six dominant eigenvalues identified earlier are considered. The plots show only the second quadrant because of symmetry about x-axis. The inter-area mode depends upon converter parameters as well as operating point of other converter. Hence, the PLL gain variation of T4 will affect the dominant pole pairs of other terminals. As the PLL gain is varied, $\lambda_1$ and $\lambda_2$ start to move towards right half plane. The point at which $\lambda_1$ and $\lambda_2$ are near to each other, the  oscillation will be generated due to controller interaction. The frequencies corresponding to controller interaction are obtained using (\ref{cond2}) and is seen to be equal to 12 Hz which correspond to gain of 35. The HVDC system becomes unstable for $K_P^{pll}\geq 35$. In Fig. \ref{RootPLL}b, the real component of  sensitivity plot of HVDC grid is depicted. The positive value of sensitivity coefficient shows the pole movement towards right half plane and vice-versa. The region in which the controller interaction takes place, the sensitivity indies close to each other. The sensitivity of $\lambda_1$ and $\lambda_2$ indicates the increase of PLL gain will reduce relative stability of HVDC grid. The sensitivity of $\lambda_2$ is high between 1 to 25 as compared to $\lambda_1$. Thus the change in $K_P^{pll}$ will repeatedly move $\lambda_2$ towards $\lambda_1$ leading to possibility of controller interaction. The conditions of controller interaction and instability derived in (\ref{cond1}) and (\ref{cond2}) are justified in the Fig. \ref{RootPLL}c. Fig. \ref{RootPLL}c shows eigenvalues and corresponding change in eigenvalues when PLL gain is changed. The point of intersection $\Delta\lambda-K_P^{pll}$ with the eigenvalue indicates the boundary point at which system is marginal stable. The system becomes unstable because the change is greater than the real component of eigenvalue. For $\lambda_1$ this point of instability is seen at $K_P^{pll} = 37$. The point at which the real component of eigenvalues cross each other or come within the distance of 4 rad/s can be identified as region of interaction. This is indicated in Fig. \ref{RootPLL}c by circle of radius equal to 4 rad/s and seen to corresponding to 15 to 25.  

\begin{figure}
\centering
    \includegraphics[width=0.5\textwidth, height=0.30\textwidth]{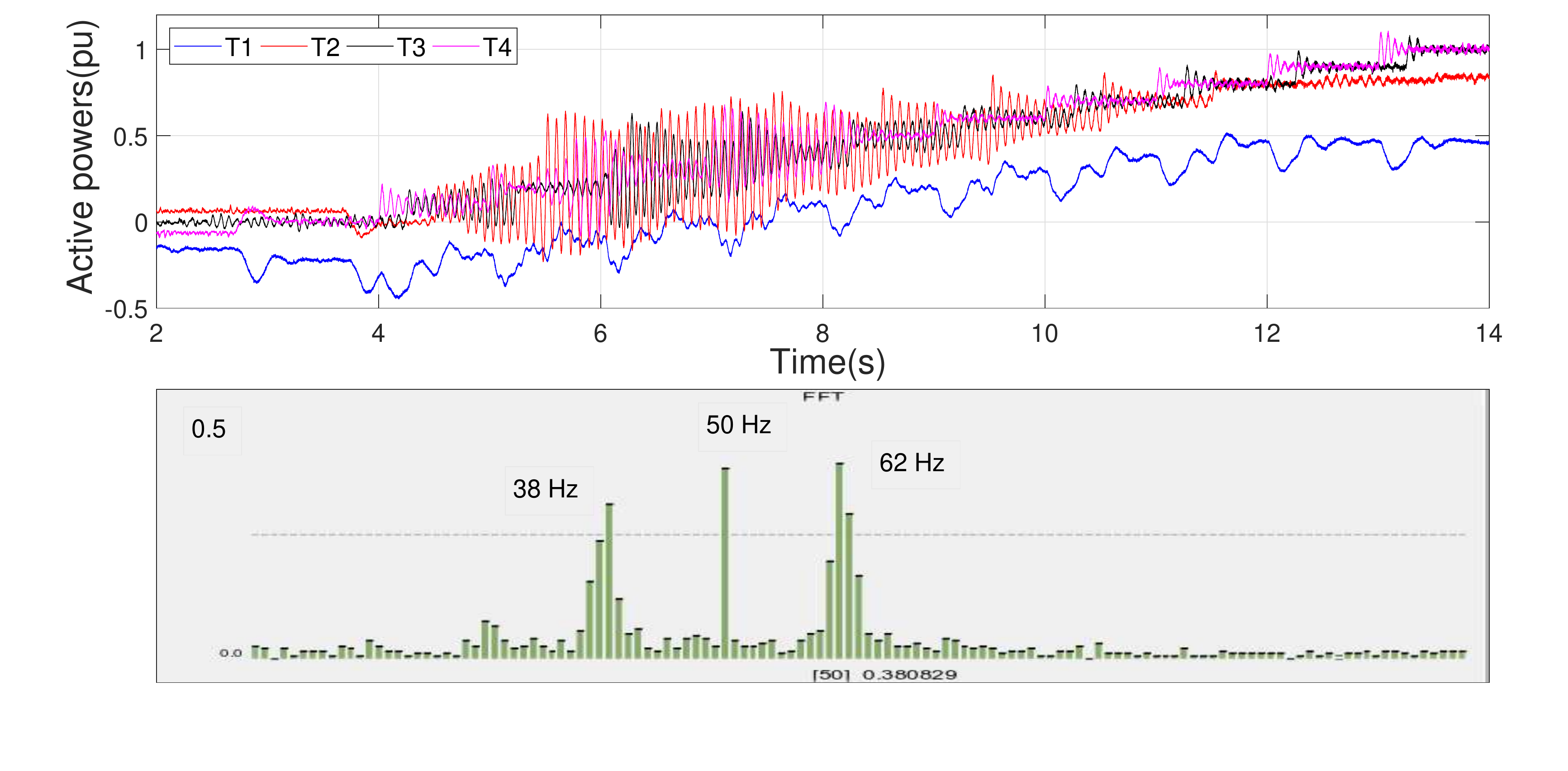}
        \vspace{-0.8cm}
    \caption{\footnotesize{HVDC grid response on Step power changes (a) Power response (b) FFT of line current}}
    \label{PowerFFT1}
    \vspace{-0.6cm}
\end{figure}

\begin{figure}
\centering
    \includegraphics[width=0.5\textwidth, height=0.30\textwidth]{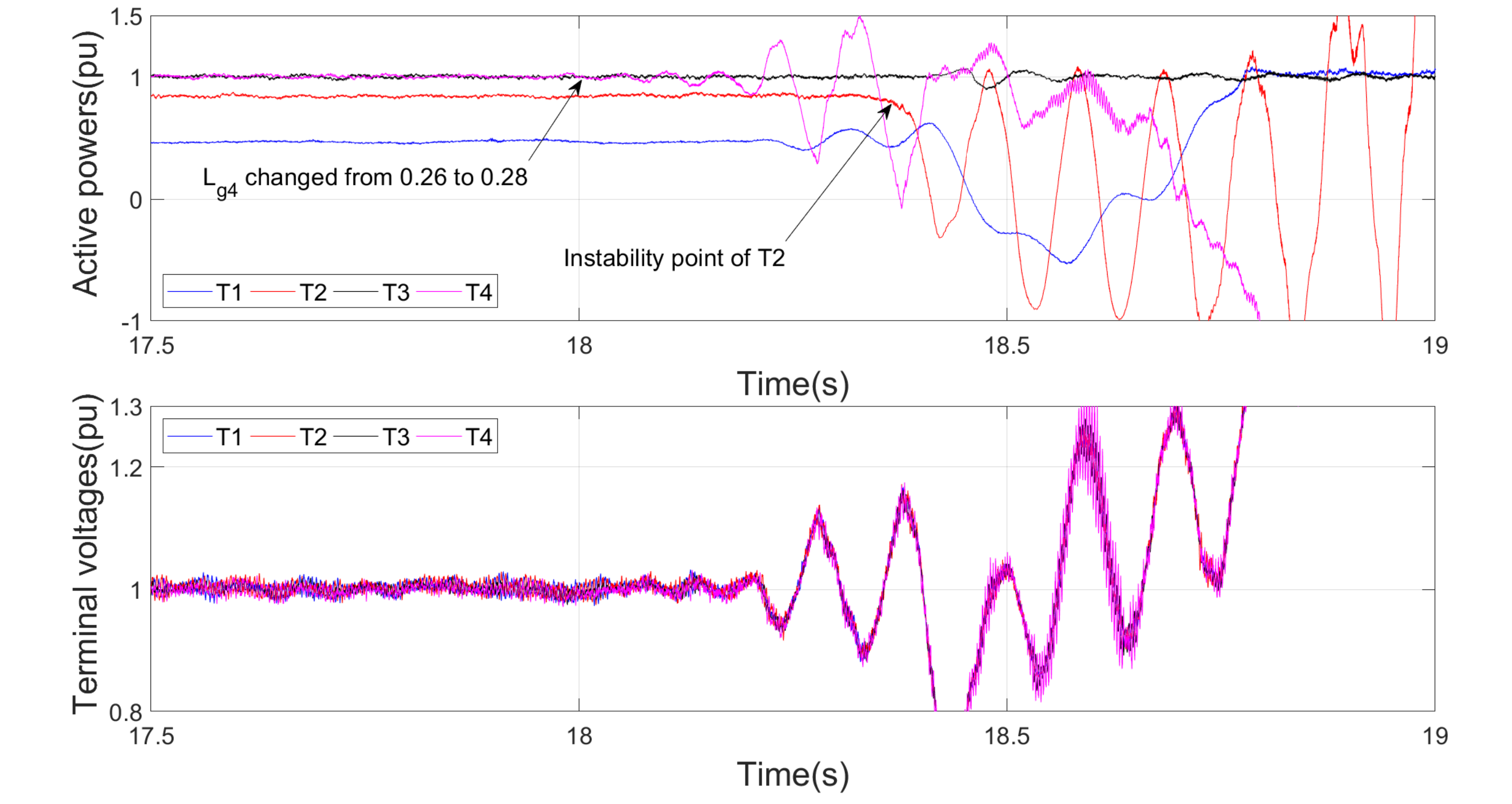}
    \caption{\footnotesize{HVDC grid response on SCR changes (a) Power response (b) Voltage response}}
    \label{PowerSCR}
    \vspace{-0.5cm}
\end{figure}

\begin{figure}[h]
\centering
    \includegraphics[width=0.5\textwidth, height=0.30\textwidth]{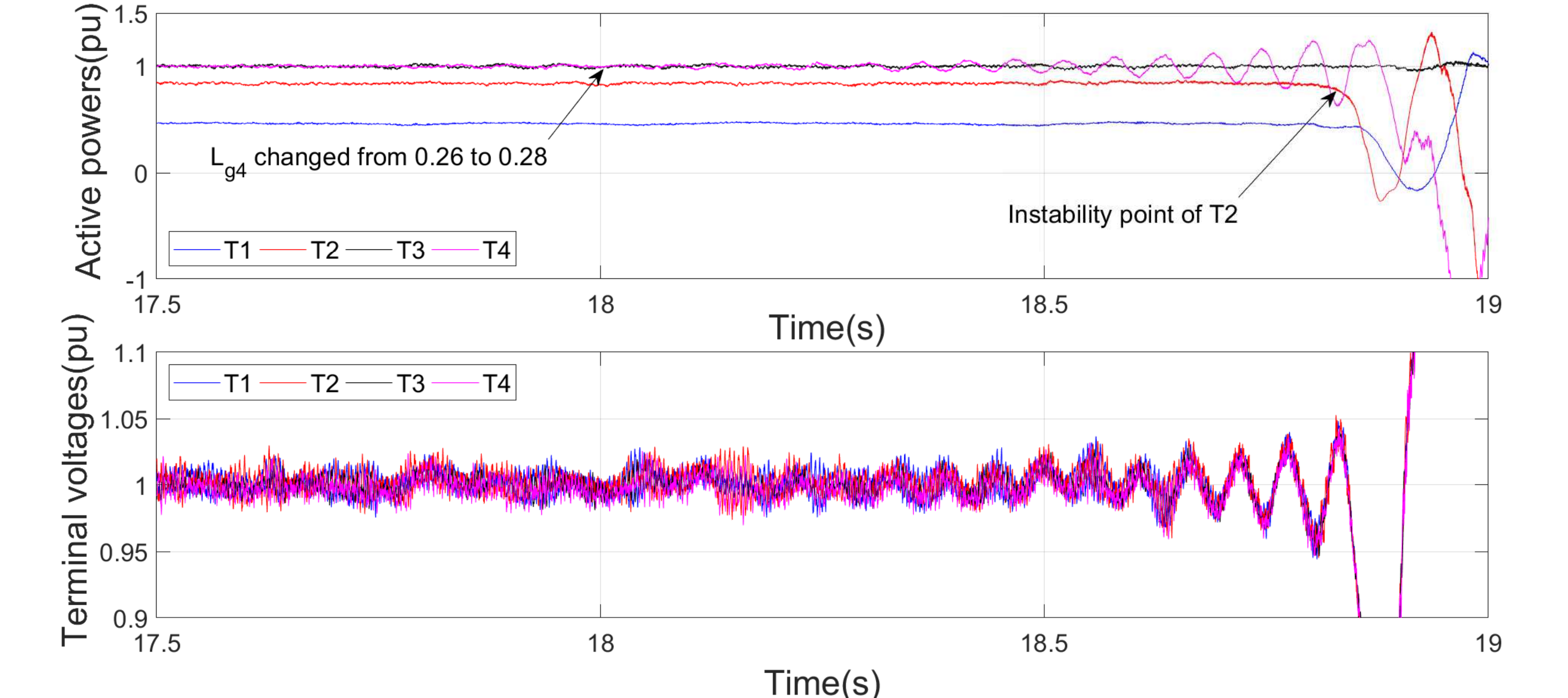}
    \caption{\footnotesize{HVDC grid response on PLL gain changes (a) Power response (b) Voltage response}}
    \label{PowerPLL}
    \vspace{-0.5cm}
\end{figure}

Similar observation is obtained when the inductance/SCR of T4 ($L_g$) is varied from 0.15 to 0.5 H.
In Fig. \ref{RootSCR}a, the root locus of HVDC grid is plotted. As the SCR is varied, the $\lambda_1$ and $\lambda_2$ start to move towards right half plane. The point at which $\lambda_1$ and $\lambda_2$ are near to each other, the oscillation will be generated due to controller interaction. The HVDC grid becomes unstable for $L_g\geq 0.29$ H. In Fig. \ref{RootSCR}b, the real component of sensitivity of HVDC grid is depicted. The positive value of sensitivity coefficient shows the pole movement toward right half plane and vice-versa. The sensitivity of $\lambda_1$ and $\lambda_2$ indicate the variation of SCR will reduce relative stability of HVDC grid. The sensitivity of $\lambda_2$ is high between 0.15 to 0.29 as compared to $\lambda_1$. Thus the change in SCR will repeatedly move $\lambda_2$ towards $\lambda_1$ leading to possibility of controller interaction.Fig. \ref{RootSCR}c shows eigenvalues and corresponding change in eigenvalues when SCR is changed. The point of intersection $\Delta\lambda-L_g$ with the eigenvalue indicates the boundary point at which system is marginal stable. The system becomes unstable because the change is greater than the real component of eigenvalue. For $\lambda_1$ this point of instability is seen at $L_g = 0.297$. The point at which the real component of eigenvalues cross each other or come within the distance of 4 rad/s can be identified as region of interaction. This is indicated in Fig. \ref{RootSCR}c by circle of radius equal to 4 rad/s and seen to corresponding to 0.26 to 0.29.
 
The switched model of HVDC grid is implemented in PSCAD. The SCR ratios and operating mode of each converter are given in the section \ref{description}. Fig. \ref{RootLoad} shows the root loci of HVDC grid when the power of T4 is varied from 0 to 1.5 pu. As the T4 power is increased the $\lambda_1$, $\lambda_4$, $\lambda_5$ and $\lambda_6$ are start to move towards right half plane and $\lambda_2$, $\lambda_3$ and $\lambda_7$ are away from imaginary axis. These conditions were replicated in the PSCAD simulation by changing the T4 power reference. Fig. \ref{PowerFFT1}a shows the change in power as obtained from PSCAD simulation. The results are seen to several frequencies and frequency spectrum is shown in Fig. \ref{PowerFFT1}b. From Fig. \ref{PowerFFT1}b, it is seen that 38 and 62 Hz are dominant frequencies which are correspond to 12 Hz when transformed to $dq$ domain. The 12 Hz frequency is also seen in Fig.  \ref{PowerSCR} and \ref{PowerPLL}. 
   
Fig. \ref{PowerSCR}a shows the power response of HVDC grid when grid inductance ($L_{g4}$) is changed from 0.26 to 0.28 H. Severe oscillation is observed on dc bus voltages. The power of T4 start to oscillates when SCR is changed however T2 which has least SCR remains stationary initially. From Fig. \ref{PowerSCR}b, it is seen that the voltage across dc grid starts to oscillate immediately at 18s. The power oscillation at T2 are triggered only when dc voltage drops to 0.8 pu. This correspond to time instant 18.4s in Fig. \ref{PowerSCR}a and \ref{PowerSCR}b. The power oscillation of T2 results in instability of HVDC grid. T3 does not show any significant oscillation even when DC grid voltage starts to deviate. This may be because of high SCR and power control mode. The simulation model becomes unstable at $L_g=$ 0.28 H whereas for small signal model instability is obtained at 0.29 H. This discrepancy due to damping factor which are not included in linearized model.   

Similar observation is obtained when the PLL gain $K_P^{pll}$ of T4 is changed from 10 to 50 at $t=18s$. From Fig. \ref{PowerPLL}a, it is seen the power of T4 starts to oscillate immediately at 18s whereas the power of T2 remain stationary initially. When the grid voltages drop to 0.8 pu at 18.8s, the power of T2 starts to oscillation. This can be seen from Fig. \ref{PowerPLL}a, and \ref{PowerPLL}b. The power oscillation of T2 results in instability of HVDC grid.     

\vspace{-0.3cm}
\section{Conclusion}
The small signal model of HVDC grid is improved for terminal interaction analysis of HVDC grid. The dc voltage dynamic is included in ac dynamic equation to see the impact of the dc grid. The local modes and inter-area modes of the HVDC grid are classified using state participation matrix. The modes transition depend upon the converter parameter, controller parameters and operating point of the converters. The participation of controller states on inter-area mode confirms the interaction between the controllers of the converters. The range of PLL gain and SCR are identified to avoid controller interaction and  stable operation of HVDC grid and it is analytically proved using the sensitivity analysis of HVDC grid. The Controllers of two terminals interacts with each other for certain value of PLL gain and SCR, therefore proper tuning of PLL gains are required. The interaction between controllers is demonstrated by varying the PLL gain and SCR of one converter. The variation of PLL gain/SCR of one converter puts limitation on other controller parameters. The weak grid connected converter is highly influenced by the parameter variation of other converter. This interaction is verified by using non-linear model of HVDC grid.      

\small
\bibliographystyle{IEEEtran}
\bibliography{IEEEexample}
\end{document}